\pdfoutput=1

\documentclass[11pt]{article}

\usepackage[final]{acl}

\usepackage{times}
\usepackage{latexsym}

\usepackage[T1]{fontenc}

\usepackage[utf8]{inputenc}

\usepackage{microtype}

\usepackage{inconsolata}

\usepackage{graphicx}
\usepackage{amssymb}
\usepackage{amsmath,cases}
\usepackage{natbib}
\usepackage{multirow}
\usepackage{subfigure}
\usepackage{epstopdf}
\usepackage{comment}
\usepackage{url}
\usepackage{xcolor}
\usepackage{bbm}
\usepackage{color}
\usepackage{soul}
\usepackage{tcolorbox}
\usepackage{array}
\usepackage{enumitem}
\usepackage{arydshln}
\usepackage{booktabs}
\usepackage{multirow}
\usepackage{amsthm,amsmath,amssymb}
\usepackage{mathrsfs}
\usepackage{multirow}
\usepackage{setspace}
\usepackage{threeparttable}
\usepackage{tabu}
\usepackage{bm}
\usepackage{pifont}

\usepackage{colortbl}

\definecolor{lightyellow}{RGB}{241,230,209}

\definecolor{y1}{RGB}{0,128,64}
\definecolor{y2}{RGB}{0,128,255}
\definecolor{y3}{RGB}{255,128,0}

\title{A Two-Stage Adaptation of Large Language Models for Text Ranking}

\author{
  Longhui Zhang$^1$ ~~~~ Yanzhao Zhang ~~~~ Dingkun Long~~~~ Pengjun Xie \\
  {\bf Meishan Zhang$^1\thanks{~~Corresponding author.}$ ~~~~ Min Zhang$^1$} \\
  $^1$Institute of Computing and Intelligence, Harbin Institute of Technology (Shenzhen)\\
    \texttt{\{longhuizhang97, zhangyanzhao00, longdingkun1993, xpjandy\}@gmail.com} \\
    \texttt{\{zhangmeishan, zhangmin2021\}@hit.edu.cn} \\
}

\begin{document}
\maketitle

\begin{abstract}
Text ranking is a critical task in information retrieval. Recent advances in pre-trained language models (PLMs), especially large language models (LLMs), present new opportunities for applying them to text ranking. 
While supervised fine-tuning (SFT) with ranking data has been widely explored to better align PLMs with text ranking goals, previous studies have focused primarily on encoder-only and encoder-decoder PLMs. Research on leveraging decoder-only LLMs for text ranking remains scarce.
An exception to this is RankLLaMA~\cite{rankllama}, which uses direct SFT to explore LLaMA’s potential for text ranking.
In this work, we propose a two-stage progressive paradigm to better adapt LLMs to text ranking.
First, we conduct continual pre-training (CPT) of LLMs on a large weakly-supervised corpus.
Second, we perform SFT, and propose an improved optimization strategy building upon RankLLaMA.
Our experimental results on multiple benchmarks show that our approach outperforms previous methods in both in-domain and out-domain scenarios.
\end{abstract}
\section{Introduction}

Text ranking is to order a set of candidate documents by their relevance to a given query. 
This process is often the second step in information retrieval, following the initial collection of candidate documents from a large corpus by a fast retriever\footnote{As the candidate set is usually small, while the first step focuses on efficiently collecting candidate documents, text ranking tends to prioritize performance over efficiency.}.
Early work relied primarily on the handcrafted numerical features based on query-document pairs~\cite{DBLP:journals/jmlr/ChapelleC11}.
Recent advances in PLMs such as BERT, along with large-scale annotated datasets like MS MARCO~\cite{msmarco}, have significantly improved model performance in text ranking.

\begin{figure}[t]
 	\centering
 	\includegraphics[width=0.48\textwidth]{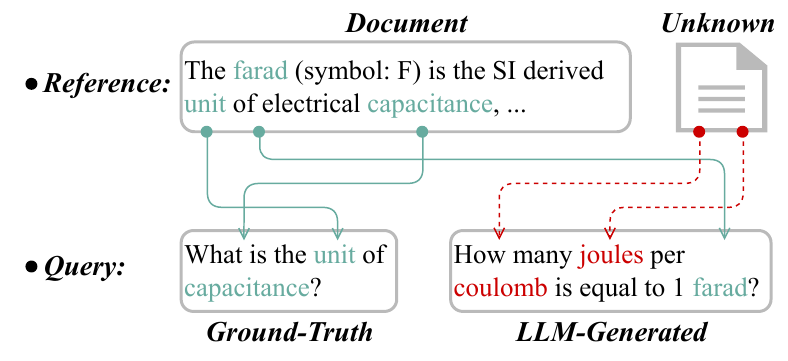}
 	\caption{Misalignment between LLMs (LLaMA) and text ranking objectives: \newcite{DBLP:conf/emnlp/SachanLJAYPZ22} measures relevance using the probability of generating a query given the document. 
Unlike ground-truth queries, LLM-generated queries could contain document-irrelevant terms. 
Such misalignment would lead to suboptimal ranking performance with out-of-the-box LLMs.}
 	\label{fig:intro}
 \end{figure}
 
LLMs, such as LLaMA~\cite{llama} and GPT4~\cite{openai2023gpt4}, have brought a paradigm shift in natural language processing through their impressive performance on various tasks.
This has driven growing interest in applying LLMs to text ranking~\cite{DBLP:journals/corr/abs-2211-09110}.
Recent works have explored prompt learning, as well as pointwise~\cite{DBLP:conf/emnlp/SachanLJAYPZ22}, pairwise~\cite{qin2023large} and listwise~\cite{sun2023chatgpt} text ranking schemas to enable out-of-the-box LLMs to perform unsupervised ranking.
With the help of LLMs, substantial improvements over the BERT-style PLM counterparts have been achieved.

However, a misalignment persists between the LLM pre-training  and the text ranking, as shown in Figure~\ref{fig:intro}.
Several studies address this through SFT of encoder-only~\cite{monobert} and encoder-decoder~\cite{rankt5} models. 
Yet rare work has targeted decoder-only LLMs.
RankLLaMA~\cite{rankllama} might be the only exceptional work exploring SFT on decoder-only LLMs. 
While RankLLaMA shows some gains, its achievements still lag behind those of previous SFT studies, and the observation could be even more serious when tested on the out-domain scenario.

In this work, we propose a two-stage training framework to adapt decoder-only LLMs to text ranking progressively:
(1) CPT followed by (2) SFT.
Given the broad definition of text relevance,
e.g., reasoning and semantic similarity, 
we first exploit a CPT stage to teach LLMs various cases of relevance. 
This helps the second-stage SFT more readily and accurately align LLMs with text ranking objectives.
During CPT, we construct a large-scale weakly-supervised text-pair dataset, and then perform the next-token prediction task (NTP)~\cite{radford2018improving} on it.
For SFT, we introduce a new optimization objective different from RankLLaMA to better explore the potential of LLMs.

We demonstrate the efficacy and generalizability of our two-stage adaptation with extensive experiments on in-domain and out-domain datasets.
We test our method on major decoder-only LLMs and various model scales, covering BLOOM 560M-7B~\cite{bloom}, LLaMA-7B~\cite{llama2}, Baichuan-7B~\cite{baichuan2}, and Qwen-7B~\cite{qwen}.
The experimental results show that our method substantially improves over its baselines, highlighting the benefits of our progressive paradigm for text ranking.
We also perform in-depth analysis into how our two-stage adaptation bridges the gap between LLMs and the text ranking task\footnote{The source code and models will be publicly available at https://github.com/Alibaba-NLP/RankingGPT.}.

\section{Method}
\subsection{Background}
Text ranking refers to the task of determining how relevant each candidate document is to a given query.
In our work, we exploit the pointwise strategy for inference, where the relevance scores are computed explicitly for each query-document pair~\cite{crammer2001pranking,monobert}.
The strategy has demonstrated high efficiency in real-world deployment compared with pairwise and listwise approaches \cite{liu2009learning}.

Formally, given a query $q$ and a set of candidate documents $D=\{d_1,\dots,d_n\}$,
we calculate relevance $\operatorname{score}(q,d_i), i \in [1,n]$ first and then execute a sorting procedure according to the scores.
We can directly use the scoring methods of the out-of-the-box LLM exploration in text ranking to obtain $\operatorname{score}(q,d_i)$, which has shown significant effectiveness~\cite{DBLP:journals/corr/abs-2211-09110,DBLP:journals/corr/abs-2308-07107}.

Here we adopt one representative scoring strategy used in \newcite{DBLP:conf/emnlp/SachanLJAYPZ22}, treating the generation probability of $q$ conditioned on $d_i$ as the relevance score:
\begin{equation}
    \label{eq:score}
    \begin{aligned}
        \mathcal{P}(d_i) &= \text{`Document: }d_i\text{ Query:'} \\
        \operatorname{score}(q,d_i)&= \prod_j p\left(q_j \mid \mathcal{P}(d_i),q_{<j}\right).
    \end{aligned}
\end{equation}
Here, $q_j$ denotes the $j$-th token of the query $q$, $q_{<j}$ represents the token sequence preceding the $j$-th token in query $q$,
and $\mathcal{P}(d_i)$ represents the document-conditioned prompt.
The calculation of each token generation probability can be parallelized, so the time complexity of this strategy is similar to that of its counterparts~\cite{monobert,monot5}.

\begin{figure}[t]
	\centering
	\includegraphics[width=0.43\textwidth]{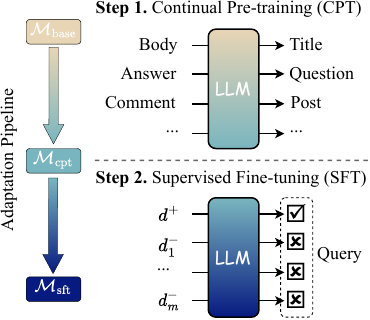}
	\caption{Two-stage adaptation paradigm. The base LLM $\mathcal{M}_{\text{base}}$ turns into an intermediate model $\mathcal{M}_{\text{cpt}}$ after CPT, and then $\mathcal{M}_{\text{cpt}}$ generates the final ranking model $\mathcal{M}_{\text{sft}}$ through SFT.}
	\label{fig:model}
\end{figure}

Although this out-of-the-box scoring method is mostly reasonable for text ranking, it would lead to suboptimal performance due to salient differences between the goals of LLM pre-training and text ranking, as illustrated in Figure~\ref{fig:intro}.
Previous studies have shown that we can better explore PLMs by adapting them with text ranking-specific training objectives~\cite{monot5,rankt5}.
Building on these studies, we propose a two-step training strategy, (1) CPT and (2) SFT, to adapt LLMs to text ranking,
as shown in Figure~\ref{fig:model}.

\begin{table*}[t]
\centering
\resizebox{\textwidth}{!}{
\begin{tabular}{lllll}
\hline
 Text Pair Format &Source  &Size & Query & Document \\
\hline
 \begin{tabular}[c]{@{}l@{}}(title, body)\end{tabular} &CommonCrawl &1.6M & \begin{tabular}[c]{@{}l@{}}\textcolor{y1}{Tango} helps support \textcolor{y3}{North Texas Food Bank}.\end{tabular} & \begin{tabular}[c]{@{}l@{}}\textcolor{y1}{Tango} Celebrates their 2013/2014 Partnership\\ with the \textcolor{y3}{North Texas Food Bank} \ldots\end{tabular} \\
\hline
\begin{tabular}[c]{@{}l@{}}(title, abstract)\end{tabular} &arXiv &1.5M & \begin{tabular}[c]{@{}l@{}}\textcolor{y1}{Duality} and \textcolor{y3}{Tameness}.\end{tabular} & \begin{tabular}[c]{@{}l@{}}We prove a \textcolor{y1}{duality} theorem and show different kinds
\\   of failure of \textcolor{y3}{tameness} of local cohomology.\end{tabular}   \\ \hline
\begin{tabular}[c]{@{}l@{}}(citation, reference)\end{tabular} &Semantic Scholar &1.2M & \begin{tabular}[c]{@{}l@{}} Some comparative growth properties of \\ \textcolor{y1}{composite entire} and \textcolor{y3}{meromorphic functions} \ldots\end{tabular} &  \begin{tabular}[c]{@{}l@{}} The aim of this paper is to prove some results\\ about \textcolor{y1}{composite entire} and \textcolor{y3}{meromorphic functions} \ldots \end{tabular} \\ \hline
(post, comment) &Reddit &1.7M & \begin{tabular}[c]{@{}l@{}}But are all \textcolor{y1}{evod 2 tanks} \textcolor{y3}{glass}? \end{tabular} &  \begin{tabular}[c]{@{}l@{}}The evod and \textcolor{y1}{evod 2 tanks} are plastic. The evod \textcolor{y3}{glass} \ldots \end{tabular} \\
\hline
\begin{tabular}[c]{@{}l@{}}(entity, description)\end{tabular} &DBPedia &0.8M & \begin{tabular}[c]{@{}l@{}}\textcolor{y1}{Economy} of \textcolor{y3}{Nigeria}. \end{tabular}   & \begin{tabular}[c]{@{}l@{}}\textcolor{y3}{Nigeria} is a middle income, mixed \textcolor{y1}{economy} and \ldots \end{tabular} \\
\hline
(question, answer) &StackExchange &1.9M & \begin{tabular}[c]{@{}l@{}}How many \textcolor{y1}{chromosomes} are in \textcolor{y3}{anaphase 2}? \end{tabular} & \begin{tabular}[c]{@{}l@{}}In \textcolor{y3}{anaphase II}, the sister chromatids present at the end of  \\ meiosis I are separated into 23 individual \textcolor{y1}{chromosomes}. \end{tabular} \\
\hline
\begin{tabular}[c]{@{}l@{}}(summary, content)\end{tabular} &CCNews &1.5M & \begin{tabular}[c]{@{}l@{}}\textcolor{y1}{Zidane} \textcolor{y2}{apologizes} for \textcolor{y3}{head butt}. \end{tabular}   &  \begin{tabular}[c]{@{}l@{}}French soccer star \textcolor{y1}{Zidane} \textcolor{y2}{apologized} for \textcolor{y3}{head-butting}  \\ an Italian opponent \ldots \end{tabular} \\ \hline
\end{tabular}
}
\caption{Examples of weakly supervised text pairs. 
Related words in queries and documents are highlighted in the same colors, showing that queries are often closely related to document content. Therefore, CPT on these data can help LLMs generate document-relevant queries, alleviating the misalignment shown in Figure~\ref{fig:intro}.}
\label{tab:weaklydata}
\end{table*}

\subsection{Continual Pre-training (CPT)}
The first stage involves CPT of LLMs on a weakly supervised relevance dataset that is automatically collected. 
As originally proposed by \newcite{DBLP:conf/acl/GururanganMSLBD20}, CPT enables task- and domain-specific adaptation of LLMs for text ranking tasks.

A straightforward strategy of adapting LLMs to text ranking tasks is to perform SFT with ranking data, and RankLLaMA~\cite{rankllama} follows this strategy.
However, obtaining such large-scale and high-quality ranking datasets would be a formidable challenge.
Moreover, large gaps between LLMs and ranking tasks may limit SFT's capability to fully explore the original knowledge in LLMs. 
A prospective solution is progressive multi-stage learning, where we first perform CPT to orient LLMs towards ranking goals before conducting final SFT for accurate alignment.

\paragraph{Weakly Supervised Data.}
Text relevance involves a range of aspects,
such as question answering, semantic similarity, summarization, description.
While LLM pre-training incorporate some of these aspects, here we further emphasize them since they are closely-related to text ranking.

To this end, we collect a large scale of text pairs covering different relevance types and domains as much as possible.
Most text pairs are sourced from public web pages, mined through tailored protocols, and filtered via normalization.
Concretely, we mine the following aspects of relevance to mock the query-document behaviors:
(title, body), (title, abstract), (citation, reference), (post, comment), (entity, description), (question, answer) and (summary, content).
Table \ref{tab:weaklydata} shows  details and examples of the weakly supervised corpus.

\paragraph{Pre-training.}
We regard shorter texts, such as titles, posts, and summaries, as queries, and their corresponding longer texts as documents.
Consistent with the typical pre-training goal of LLMs, we employ the NTP task on weakly supervised text pairs.
The loss function $\mathcal{L}_{\text{ntp}}(q,d)$ is as follows:
\begin{equation}
\mathcal{L}_{\text{ntp}}(q,d) = - \sum_j \log p\left(q_j \mid \mathcal{P}(d),q_{<j}\right),
\label{eq:pretrain_loss}
\end{equation}
which is equivalent to the log-likelihood of the relevance score defined in Eq.\ref{eq:score}.

\subsection{Supervised Fine-tuning (SFT)}
\label{sec:sft}
The second stage of our method is SFT, which helps further align the LLM for text ranking.
SFT has been a common technique to accurately adapt LLMs for specific tasks~\cite{rankllama}.
Here we describe the supervised data first and then introduce the objectives for effective fine-tuning.

\paragraph{Supervised Training Data.}
We leverage the MS MARCO dataset for SFT, which comprises 8.8 million documents and 53,000 positive query-document pairs.
Almost all positive text pairs in MS MARCO have been manually annotated, so this dataset is often used to train ranking models~\cite{monobert,monot5}.
In our work, we first employ the BGE model~\cite{bge_embedding} to retrieve the top 1000 negative document candidates (i.e., the most relevant documents) for each query. 
Following this, we construct the training dataset by randomly sampling corresponding positive and negative documents from the retrieved candidates.

\paragraph{The Ranking Objective.}
As demonstrated in previous text ranking studies~\cite{rankllama,rankt5},
the ranking loss~\cite{DBLP:conf/icml/ChenK0H20} based on a query $q$ and the associated list of positive and negative documents $D=\{d^+,d_1^-,\dots,d_m^-\}$ can effectively align LLMs with ranking tasks.
The ranking loss is formulated as:
\begin{equation}
\mathcal{L}_{\text{rank}}(q, D)=-\log \frac{\exp(\mathcal{S}(q, d^+)/\tau)}{\sum_{d \in D} \exp(\mathcal{S}(q, d)/\tau)},
\label{rankloss}
\end{equation}
where $\tau$ denotes the temperature parameter, and $\mathcal{S}(.)$ is the relevance scoring function.

Our objective differs from that of \newcite{rankllama} as we use $\operatorname{score}(.)$ in Eq.~\ref{eq:score} as the scoring function $\mathcal{S}(.)$ rather than exploiting the last-token representation for relevance scoring.
$\operatorname{score}(.)$ incorporates more scoring evidence by considering all query tokens.
More importantly, our objective is to directly optimize NTP probabilities, a key property of LLMs, to fit the ranking goal.
In this way, we can benefit from the strong generalization capabilities of pre-trained LLMs.

\paragraph{Auxiliary Objectives.}
LLMs have potential for better handling out-domain scenarios as they are pre-trained on a large and diverse corpora~\cite{openai2023gpt4}.
Yet the large-scale parameters of LLMs might cause overfitting to the training dataset.
To avoid this problem, we supplement the ranking optimization with two additional objectives.
The first one is the NTP objective $\mathcal{L}_\text{ntp}$, which is consistent with CPT and utilizes positive text pairs in SFT data.
The second one is newly designed by us, namely Differential Penalty (DP) as follows:
\begin{equation}
   \mathcal{L}_\text{dp}(\mathcal{M}_{\text{cpt}},\mathcal{M}_{\text{sft}})=\frac{1}{\lVert T \rVert}\sum\limits_{j}^{\lVert T \rVert}\sum\limits_{k}^{\lVert V \rVert}\operatorname{KL}(p_{\text{cpt}}^{j,k},p_{\text{sft}}^{j,k}),
			\label{kl}
\end{equation}
where $\operatorname{KL}$ is the Kullback-Leibler (KL) divergence, $V$ is the model vocabulary, and $T$ is all query tokens.
$p_{\text{cpt}}^{j,k}$ and $p_{\text{sft}}^{j,k}$ denote the token probabilities calculated by the model $\mathcal{M}_{\text{cpt}}$ and the model $\mathcal{M}_{\text{sft}}$ respectively.
The DP objective actually constrains the generation difference between the adapted model and the initialized model.

Overall, our mixed objective loss function during SFT is as follows:
\begin{equation}
 \mathcal{L}_{\text{sft}} = \alpha \mathcal{L}_\text{rank}+ (1-\alpha) (\mathcal{L}_\text{ntp}+\mathcal{L}_{\text{dp}}),
    \label{eq:overall}
\end{equation}
where $\alpha$ is the trade-off hyper-parameter.

	\begin{table*}[t]
		\centering
		\small
		\begin{tabular}{cccccccccc}
			\toprule
			\multirow{3}{*}{Method} &\multirow{3}{*}{LLM} &\multirow{3}{*}{Size} & \multicolumn{3}{c}{Sparse Retrieval - BM25} & \multicolumn{3}{c}{Dense Retrieval - BGE} & \multirow{3}{*}{Average} \\
			\cmidrule(r){4-6} \cmidrule(r){7-9}
			& & &MS MARCO &DL19 &DL20 &MS MARCO &DL19 &DL20    \\ \midrule
                    Retrieval &NA &NA &22.8 &50.6 &48.0 &40.9 &71.4 &70.5 &50.7\\\hdashline
                    MonoBERT &BERT &340M &44.0 &72.3 &70.3 &44.7 &72.0 &70.2 &62.3\\\hdashline
                    \multirow{3}{*}{MonoT5} &T5 &220M &43.6 &71.5 &69.7 &43.4 &69.4 &65.8 &60.6\\
                     &T5 &770M &43.4 &73.2 &71.2 &43.5 &72.0 &70.1 &62.2\\
                     &T5 &3B &44.9 &72.8 &74.5 &45.7 &72.5 &74.5 &64.2\\\hdashline
                    RankLLaMA &LLaMA &7B &46.9	&74.4	&\bf 76.4	&47.9	&74.7	&76.2 & 66.1\\\midrule
                    \multirow{7}{*}{\colorbox{lightyellow}{{TSARankLLM}}} &BLOOM &560M &44.0 &75.3 &73.2 &44.8 &75.0 &73.7 &64.3\\
                     &BLOOM &1B &44.5 &75.6 &72.3 &45.4 &75.4 &72.9 \bf &64.4\\
                     &BLOOM &3B &45.1 &76.8 &73.6 &45.9 &76.2 &74.4 &65.3\\
                     &BLOOM &7B &46.0 &\bf 77.3 &74.6 &47.0 &\bf 77.1 &75.9 &66.3\\
                     &LLaMA &7B &46.6 &76.2 &76.3 &47.7 &76.7 &\bf 76.8 &\bf 66.7\\
                     &Baichuan &7B &46.6 &75.9 &74.3 &47.7 &75.2 &76.2 &66.0\\
                     &Qwen &7B &\bf 48.0 &75.8 &74.3 &\bf 49.0 &75.5 &75.0 &66.3\\
			\bottomrule
		\end{tabular}
		\caption{In-domain results of various models.}
		\label{tab:indomain} 
	\end{table*}

\section{Experiments}

\subsection{Experimental Settings}
\paragraph{Test Datasets.}
We use the same experimental settings as \newcite{rankllama}, covering both in-domain and out-domain scenarios.

For the in-domain scenario, we test on MS MARCO \cite{msmarco}, DL19~\cite{DBLP:journals/corr/abs-2003-07820} and DL20~\cite{DBLP:journals/corr/abs-2102-07662} benchmarks, and construct candidate documents based on the top 1000 documents retrieved by BM25~\cite{DBLP:journals/ftir/RobertsonZ09} and the top 200 documents retrieved by BGE~\cite{bge_embedding} respectively.

For the out-domain scenario, we test on BEIR benchmark \cite{beir} and use the top 1000 documents retrieved by BM25 as candidate documents.
The BEIR benchmark covers a variety of domains and ranking tasks, and therefore could be used to measure the generalization ability of ranking models.

\paragraph{Implementation Details.}
We train the model on 8 NVIDIA A100 GPUs with 80GB of memory.
During CPT, we train for 1 epoch on all weakly supervised data.
During SFT, we train for 1 epoch on the MS MARCO training set.
Following Eq.~\ref{rankloss}, we set the number of negative examples $m$ to 48 and the temperature parameter $\tau$ to 0.001.
The trade-off hyper-parameter $\alpha$ in Eq.~\ref{eq:overall} is set to 0.6.
Similar to previous work~\cite{rankllama}, we fine-tune the top 16 transformer layers and freeze other parameters to reduce GPU memory of SFT.

\paragraph{Baselines and Metric.}
We compare text ranking models with different structures in previous works, including encoder-only MonoBERT~\cite{monobert}, encoder-decoder MonoT5~\cite{monot5} and RankT5~\cite{rankt5}, and decoder-only RankLLaMA~\cite{rankllama}. Following standard practice, we adopt NDCG@10 as the evaluation metric.

\subsection{Main Results} \label{main}

 \begin{table*}[ht]
	\centering
\small
	\begin{tabular}{ccccccc|ccc}
		\toprule
		\multirow{4}{*}{Dataset} &\multicolumn{1}{c}{BM25} & \multicolumn{1}{c}{ MonoBERT} & \multicolumn{3}{c}{MonoT5} & \multicolumn{1}{c}{ RankT5}  & \multicolumn{3}{c}{\colorbox{lightyellow}{{TSARankLLM}}} \\
		\cmidrule(r){2-2} \cmidrule(r){3-3} \cmidrule(r){4-6} \cmidrule(r){7-7} \cmidrule(r){8-10}
		&\multicolumn{1}{c}{NA} & \multicolumn{1}{c}{BERT} & \multicolumn{3}{c}{T5} & \multicolumn{1}{c}{T5}  & \multicolumn{3}{c}{BLOOM} \\
		&NA &340M &220M &770M &3B &770M &560M &1B &3B      \\ \midrule
		Arguana &39.7 &51.5 &13.2  &30.2 &28.8 &33.0 &53.3 &55.1& \bf 55.6 \\
		Climate &16.5 &24.9 &24.5 &25.9 &\bf 28.0 &21.5 &22.3 &23.6 &27.7\\
		DBPedia &31.8 &43.5 &42.0 &43.5 &47.8&44.2 &44.5 &45.4 &\bf 50.0 \\
		FEVER &65.1  &81.3 &80.2 &82.8 &\bf 85.0&83.2 &83.6 &82.3 &83.7 \\
		FiQA &23.6 &36.8 &41.4 &44.6 &\bf 51.4&44.5 &40.0 &42.0 &44.9 \\
		HotpotQA &63.3 &73.5 &69.5 &73.6 &75.9 &71.0 &75.6 &75.4 &\bf 76.4\\
		NFCorpus &33.8 &36.9 &35.7 &38.4 &38.4 &38.1 &37.3 &37.9 &\bf 39.4\\
		NQ &30.6 &56.8 &56.7 &60.8 &\bf 63.3&61.4 &56.1 &57.6 &58.7\\
		Quora &78.9 &71.5 &82.3 &\bf 85.4&84.1 &83.1 &82.9 &82.3 &82.9 \\
		SCIDOCS &14.9 &15.7 &16.5 &19.1 &\bf 19.7 &18.1 &18.1 &18.7 & 19.2\\
		SciFact &67.9 &72.0 &73.6 &75.5 &77.7 &75.0 &77.1 &76.3 &\bf 78.0\\
		COVID &59.5 &65.0 &77.8 &82.3 &79.5 &80.7 &78.9 &81.7 &\bf 83.3\\
		Touche &\bf 44.2 &27.7 &27.7 &28.5 &30.0 &44.0&28.2 &29.7 &30.2 \\ \midrule
		Average &48.3 &50.5 &49.3 &53.1 &54.6 &53.7 &53.7 &54.5 &\bf 56.2 \\
		 \bottomrule
	\end{tabular}
	\caption{Out-domain results of 220M-3B models.}
	\label{tab:outdomain}
\end{table*}

To verify the broad effectiveness of our method, we compare its performance with other adaptation approaches on four foundation LLMs of different types and sizes: BLOOM 560M-7B~\cite{bloom}, LLaMA-7B~\cite{llama2}, Qwen-7B~\cite{qwen} and Baichuan-7B~\cite{baichuan2}.
Our \textbf{t}wo-\textbf{s}tage \textbf{a}daptation of \textbf{LLM}s for text \textbf{rank}ing is denoted as TSARankLLM.

\paragraph{In-Domain Evaluation.}
Table~\ref{tab:indomain} summarizes the in-domain performance of our models as well as several previous works.
We find that decoder-only LLMs exhibit substantial ranking capabilities.
The proposed TSARankLLM models outperform MonoT5 models of similar scale in almost all cases.
Aligning with prior works \cite{monot5}, we observe ranking performance generally increase with model size. 
For example, increasing the size of BLOOM from 560M to 7B improves the average NDCG@10 score by $66.3 - 64.3 = 2.0$. 

Following, we look into our TSARankLLM models in terms of different foundation LLMs.
By comparing these models of the same 7B-scale parameter size,
we can see that LLaMA-7B achieves the best performance overall, surpassing BLOOM, Baichuan and Qwen.
While other foundation LLMs occasionally surpass LLaMA-7B on certain datasets (e.g. BLOOM-7B on DL19 and Qwen-7B on MS MARCO), LLaMA-7B appears to be the most effective foundation LLM of those examined.

Finally, we compare TSARankLLM and RankLLaMA adaptations on the same LLaMA-7B foundation.
As shown, our two-stage adaptation approach surpasses the single-stage SFT of RankLLaMA, yielding average improvements of $66.7-66.1=0.6$ NDCG@10.
Overall, our TSARankLLM method can provide state-of-the-art performance for the in-domain setting.

\paragraph{Out-Domain Evaluation.}
Table~\ref{tab:outdomain} shows the results of models within 3B scale, while Table~\ref{tab:outdomain-7b} provides the results of 7B models.
In general, out-domain performance significantly lags behind that of the in-domain, with a gap around 10 points, indicating that out-domain text ranking is challenging.

After examining the results in Table~\ref{tab:outdomain}, we find that the overall tendency is consistent with that of in-domain results.
Increased model size generally improves performance.
Notably, our TSARankLLM model based on BLOOM-3B outperforms the monoT5-3B of the same model size, demonstrating its strong text ranking capability.

Further, we analyze the out-domain results of 7B-scale models in Table~\ref{tab:outdomain-7b}.
We can see that our TSARankLLM models exhibit much better performance than RankLLaMA.
With the same LLaMA-7B as backend, our method gets higher NDCG@10 scores on all the datasets,
 and the averaged increase reaches $57.8-52.5=5.3$.
Among the foundation models, Qwen achieves the highest average score with our TSARankLLM approach, while LLaMA excels on most out-domain datasets.
Overall, all the results indicate that LLMs provide a promising solution for out-domain text ranking if pre-trained LLM knowledge can be effectively explored.

\begin{table}[t]
	\centering
	\small
	\begin{tabular}{ccccccc}
		\toprule
		\multirow{2}{*}{Dataset}  &RankLLa. & \multicolumn{4}{c}{\colorbox{lightyellow}{{TSARankLLM}}} \\
		\cmidrule(r){2-2} \cmidrule(r){3-6}
		  &LLa. &{BLO.} & {LLa.} &{Bai.} &{Qwen} \\ \midrule
		Arguana &47.0 &56.1 &51.2 &54.1 & \bf 56.8 \\
		Climate &19.1 &28.1 &31.6 &32.2 &\bf 37.2 \\
		DBPedia&48.6 &47.8 &49.2 &\bf 50.0 &48.5 \\
		FEVER &74.5 &84.1 &84.5 &83.0 &\bf 85.6 \\
		FiQA &42.2 &46.4 &\bf  50.2 &49.4 &49.6\\
		HotpotQA &75.2 &77.4 &\bf 80.7 &80.0 &80.4\\
		NFCorpus &35.8 &39.5 &40.2 &39.8 &\bf 40.4\\
		NQ &62.1 &60.4&\bf 63.6 &63.1 &62.7\\
		Quora &80.5 &83.8&\bf 85.9 &85.0 &82.6\\
		SCIDOCS &19.0 &\bf 20.3 &19.8 &20.0 &19.8\\
		SciFact &70.1 &77.8 &\bf 78.3 &77.6 &77.6\\
		COVID &77.4 &82.9 &\bf 84.0 &82.4 &83.7\\
		Touche &31.0 &31.0 &32.0 &30.6 &\bf 32.6\\ \midrule
		Average &52.5 &56.6 &57.8 &57.5 &\bf 58.3\\
		\bottomrule
	\end{tabular}
	\caption{Out-domain results of 7B models. ``RankLLa.'', ``BLO.'', ``LLa.'' and ``Bai.'' represent RankLLaMA, BLOOM, LLaMA and Baichuan respectively.}
	\label{tab:outdomain-7b} 
\end{table}

\subsection{Ablation Analysis} \label{ablation}

In this subsection, our ablation analysis quantifies the contribution of each part of our two-stage adaptation to the overall performance improvements of the TSARankLLM model.

\paragraph{The Two-Stage Training.}
Table~\ref{tab_ablation} shows the individual contributions of CPT and SFT based on BLOOM-560M and LLaMA-7B.
As shown, both training stages exhibit significant performance gains,
validating their importance to optimal final results.
In particular, the influence of SFT is highly remarkable.
This is unsurprising given its use of high-quality training data from MS MARCO to accurately align LLMs with ranking objectives.
Without both stages, our models degenerate to out-of-the-box LLMs with unsatisfactory ranking capabilities, whose in-domain performance can be even worse than direct retrieval without ranking.
This directly indicates that the misalignment between out-of-the-box LLMs and text ranking objectives results in suboptimal performance.

\begin{table}[t]
	\centering
        \small
	\begin{tabular}{lcccc}
		\toprule
		 \multicolumn{1}{c}{\multirow{3}{*}{Method}} & \multicolumn{2}{c}{BLOOM-560M} & \multicolumn{2}{c}{LLaMA-7B} \\
		\cmidrule(r){2-3} \cmidrule(r){4-5}
		&In. &Out. &In. &Out.   \\ \midrule
 \colorbox{lightyellow}{{TSARankLLM}} &\bf 64.3 &\bf 53.7 & \bf 66.7 &\bf 57.8 \\ \midrule
  \multicolumn{5}{l}{$\bullet $\emph{{ Two-Stage Training}}} \\
		\quad - CPT &63.8 &51.0 &66.3 &55.7 \\
	\quad- SFT &53.6 &47.8 &56.3 &52.6 \\
        \quad - CPT\&SFT  &43.4 &43.4 &47.5 &48.6 \\ \hdashline
		\multicolumn{5}{l}{$\bullet $\emph{ Auxiliary Objectives of SFT}} \\
        \quad- $\mathcal{L}_\text{ntp}$ &\bf 64.3 &53.2 &66.6 &57.0 \\
        \quad- $\mathcal{L}_{\text{dp}}$ &64.2 &52.7 &66.4 &56.3 \\
        \quad - $\mathcal{L}_\text{ntp}$\&$\mathcal{L}_{\text{dp}}$ &64.2 &52.0 &66.4 &55.5 \\
		\bottomrule
	\end{tabular}
	\caption{Ablation results of BLOOM-560M and LLaMA-7B in in-domain (In.) and out-domain (Out.) scenarios.}
	\label{tab_ablation} 
\end{table}

\begin{figure}[t]
	\centering
	\includegraphics[width=0.35\textwidth]{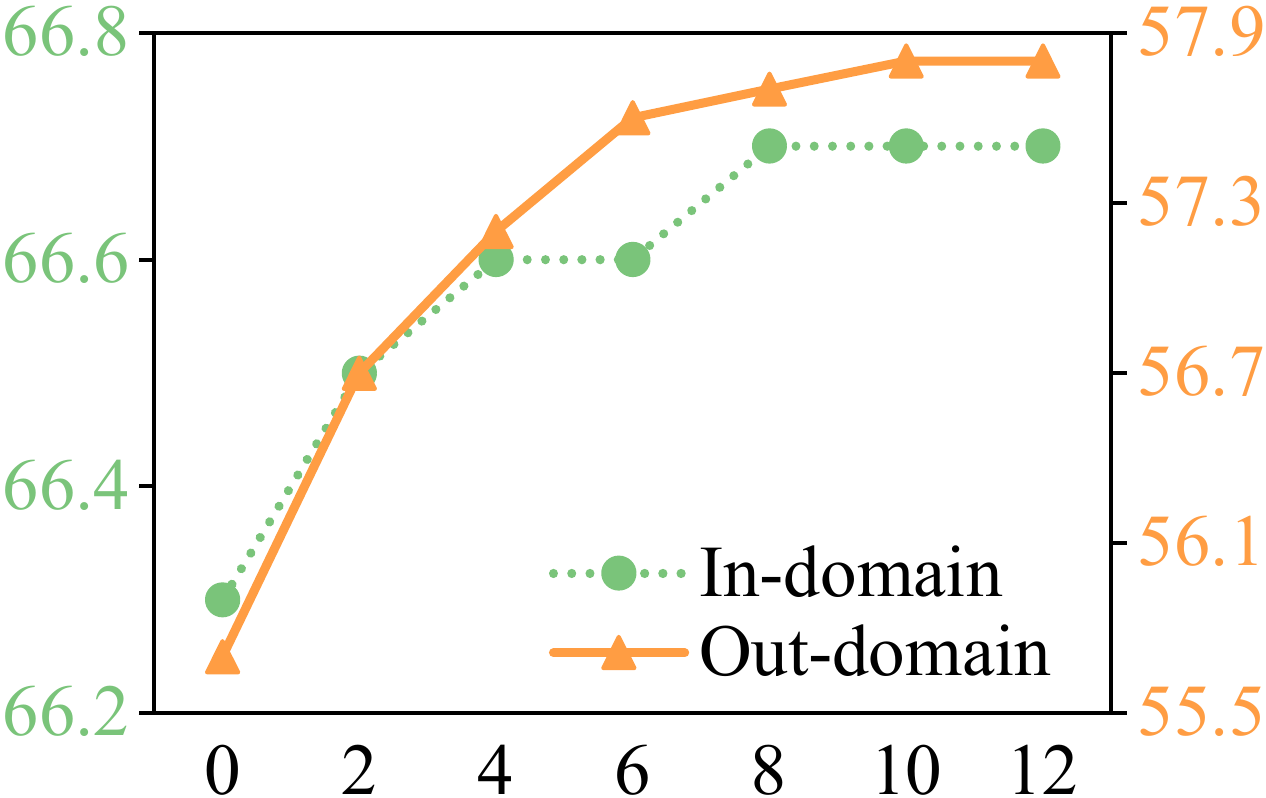}
	\caption{Results of TSARankLLM based on LLaMA-7B at various CPT data sizes (millions).}
	\label{cpt_scale} 
\end{figure}

\paragraph{The Scale of Weakly-Supervised Data in CPT.}
During CPT, we construct a weakly-supervised dataset.
An important question is that how the data scale influences our model performance. 
Figure \ref{cpt_scale} shows the results of our method on LLaMA-7B, illustrating trends for both in-domain and out-domain performance.
As expected, model performance improves with greater data scale, though the gains become insignificant after the data scale surpasses 10M.
Additionally, we can see that CPT can benefit the out-domain performance more than the in-domain setting,
as evidenced by the steeper curve of the out-domain case.

\paragraph{The Ranking Objective of SFT.}
\begin{figure}[t]
	\centering
	\includegraphics[width=0.48\textwidth]{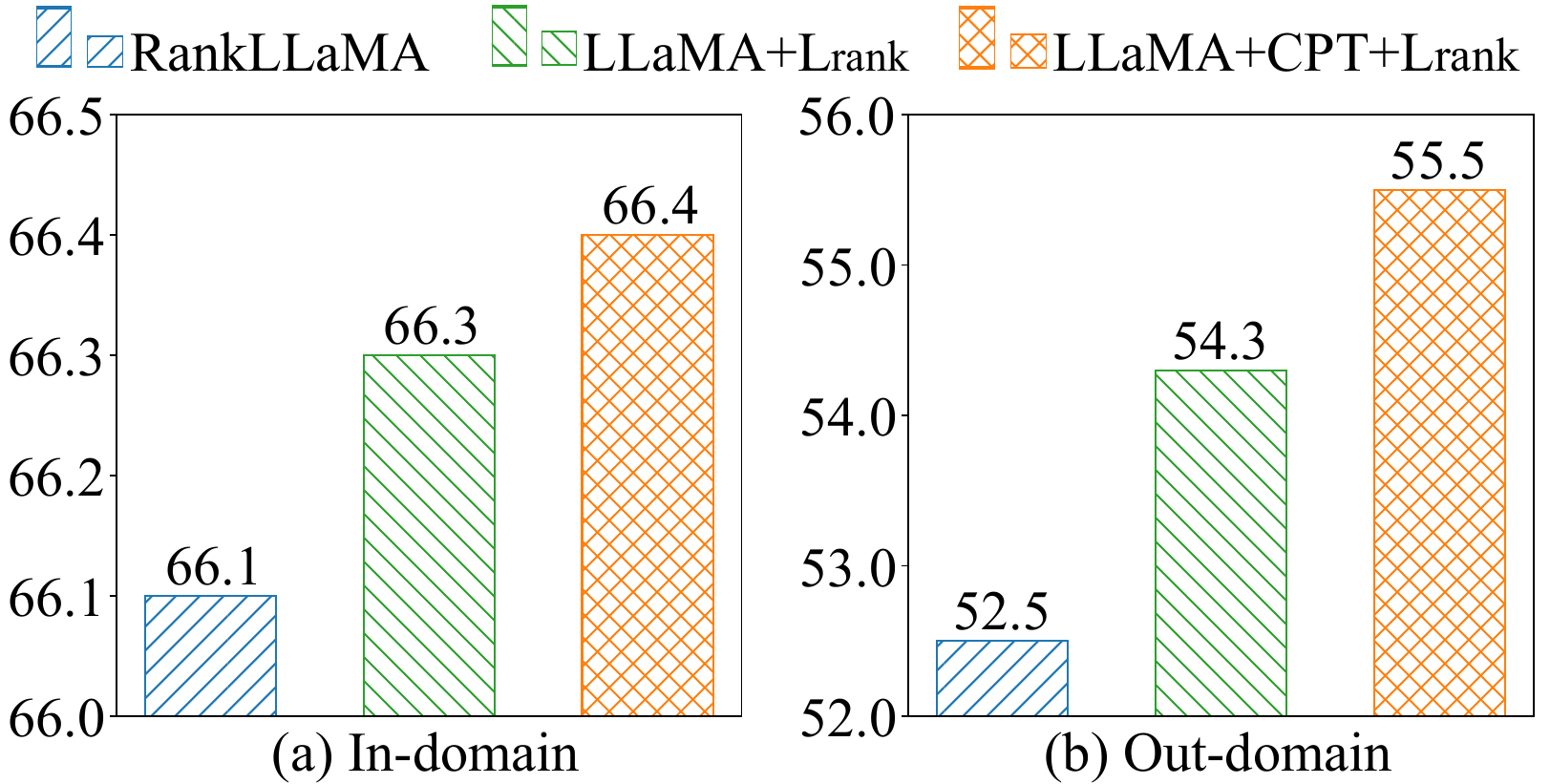}
	\caption{Model results with different ranking objectives.
RankLLaMA is based on the last token, while two variants of our models are based on entire query tokens and do not involve our two auxiliary objectives.}
	\label{fig:rankllama} 
\end{figure}
As mentioned in Section \ref{sec:sft}, we exploit a ranking objective different from that of RankLLaMA.
Here we fairly compare both ranking objectives in two settings: with and without CPT.
Both settings during SFT only exploit the ranking objective, i.e., the auxiliary objectives are removed. 
Figure~\ref{fig:rankllama} shows that our ranking objective substantially outperforms that of RankLLaMA in both settings in terms of either in-domain or out-domain performance.
The results indicate that our SFT can better align LLM pre-training with the ranking goal by optimizing the full-query generation probabilities directly.

\paragraph{The Auxiliary Objectives of SFT.}
To avoid overfitting and better explore the potential of LLMs,
we design two auxiliary objectives (i.e., $\mathcal{L}_\text{ntp}$ and $\mathcal{L}_{\text{dp}}$ ) during SFT.
Table~\ref{tab_ablation} conducts ablation analysis to test the effectiveness of the two objectives based on BLOOM-560M and LLaMA-7B.
We can see that the two objectives have negligible impact on in-domain performance, with a maximum drop of only 0.3 point when removed.
As expected, they significantly impact the out-domain performance, decreasing performance by $\frac{(53.7-52.0)+(57.8-55.5)}{2} = 2.0$ points on average when omitted. 
Notably, removing $\mathcal{L}_{\text{dp}}$ leads to a more substantial decrease, indicating its greater impact.
Moreover, we find that LLaMA-7B is more sensitive to the auxiliary objectives than BLOOM-560M,
probably because larger models can be more easily overfitted.

\subsection{Discussion}
In this subsection, we conduct detailed experimental analyses to comprehensively evaluate our two-stage training method.

\paragraph{A Comparison with More Powerful LLMs.}
Compared to the LLMs investigated in our work, some recent LLMs like UL2-20B~\cite{tay2023ul2}, ChatGPT~\cite{ChatGPT} and GPT-4~\cite{openai2023gpt4} have shown more superior performance on many NLP tasks owing to their extremely large model sizes and pre-training data.
We apply our TSARankLLM method to train the LLaMA-7B, and compare its performance against these LLMs. 
Due to the massive model sizes and even closed-source nature of these LLMs, directly training them is challenging.
Here we apply out-of-the-box ranking strategies to them: 
(1) the pairwise ranking strategy of \newcite{qin2023large} based on UL2-20B and (2) the listwise ranking strategy of \newcite{sun2023chatgpt} based on ChatGPT and GPT4.
Table~\ref{tab:uns-bm25} shows the results on five out-domain datasets selected for this comparison\footnote{The in-domain evaluation is unfair for out-of-the-box ranking strategies.}.
We can see that our method is very competitive, with an average gap of only $56.9-56.7=0.2$ point compared to GPT-4.
Considering the higher costs of these compared systems, our method could be preferable in practice.

\begin{table}[t]
	\centering
	\resizebox{0.47\textwidth}{!}{
	\begin{tabular}{c|cccc}
		\toprule
		\multirow{2}{*}{Dataset}  &Pairwise &\multicolumn{2}{c}{Listwise} & \colorbox{lightyellow}{{TSARankLLM}} \\
		\cmidrule(r){2-2} \cmidrule(r){3-4} \cmidrule(r){5-5}
		  &UL2-20B &{ChatGPT} & {GPT-4} &{LLaMA-7B}  \\ \midrule
		COVID  & 79.5	&76.7	&\bf 85.5	&84.0 \\
            NFCorpus	&36.1	&35.6	&38.5	&\bf 40.2 \\
            Touche	&37.9	&36.2	&\bf 38.6	&32.0 \\
            DBPedia	&46.5	&44.5	&47.1	&\bf 49.2 \\
            SciFact	&73.3	&70.4	&75.0	&\bf 78.3  \\ \midrule
            Average	&54.7	&52.7	&\bf 56.9	&56.7 \\
		\bottomrule
	\end{tabular}}
 	\caption{Comparison with more powerful LLMs. Due to high time complexity and API call costs of pairwise and listwise ranking strategies~\cite{qin2023large}, we only experiment with small datasets.}
	\label{tab:uns-bm25}
\end{table}

\begin{table}[t]
	\centering
        \small
	\begin{tabular}{lcc}
		\toprule
		 \multicolumn{1}{c}{Method} & Doc-word $\uparrow$ & Stop-word $\downarrow$ \\ \midrule
           Base LLaMA-7B &25.1	&19.7 \\
           \quad+ CPT &27.8	&16.1 \\
           \quad+ SFT &28.3	&15.7 \\
            \quad+ CPT\&SFT &31.7	& 14.1 \\ \midrule
            GPT-4  &\bf 33.6   & 13.6  \\
            Ground-Truth  &33.3 & \bf 13.3  \\
		\bottomrule
	\end{tabular}
	\caption{The proportion of document-relevant and stop words in queries generated by various models. 
 ``Ground-Truth'' denotes manually annotated positive queries.}
	\label{tab:token-ana} 
\end{table}

\paragraph{Quality Assessment of Query-Generation in Our Two-Stage Training.}
Query generation is a key module in our two-stage training. 
To understand this module, we conduct a human evaluation to measure the quality of query-generation directly. 
As mentioned in Figure \ref{fig:intro}, without text-ranking-oriented alignment, 
the out-of-the-box LLMs commonly generate queries including a large proportion of document-irrelevant information.  
As such, we mainly measure the quality by the percentage of document-relevant semantically-equivalent words,
while also considering adverse stop words.    
Table \ref{tab:token-ana} shows the manual evaluation results on 500 randomly selected BEIR samples.
The results of GPT4 and ground-truth are also provided for reference.
We can see that by applying CPT and SFT gradually, 
the percentage of document-relevant words increases while that of stop words decreases, both approaching the percentages of GPT4 and ground-truth.
However, higher percentage of document-relevant words and lower percentage of stop words do not necessarily indicate better performance.

\begin{figure}[t]
	\centering
	\includegraphics[width=0.40\textwidth]{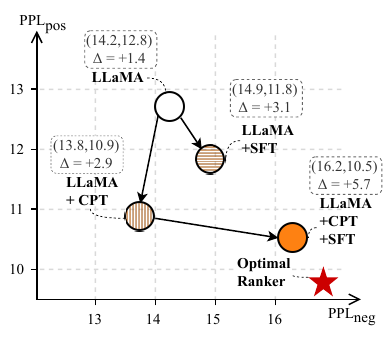}
	\caption{The perplexity (PPL) comparison of generating positive queries for a given document. ``$\Delta$'' denotes $\text{PPL}_\text{neg}-\text{PPL}_\text{pos}$, which roughly indicates the model's ranking capability.}
	\label{fig:ppl}
\end{figure}

\paragraph{A Perplexity Perspective to Examine Our Two-Stage Training.}
An optimal ranker based on Eq.~\ref{eq:score} should strongly prefer generating positive over negative queries for a given document.
Therefore, the perplexity (denoted as PPL) difference between negative and positive queries can roughly estimate the ranking capability of our models.
We randomly select 5,000 positive and negative text pairs from BEIR and test the query-generation PPL of various rankers, as shown in Figure~\ref{fig:ppl}.
We observe that in our two-stage training pipeline, CPT significantly reduces the PPL of positive queries (i.e, $\text{PPL}_\text{pos}$),
and SFT increases PPL of negative queries (i.e., $\text{PPL}_\text{neg}$) remarkably. 
The observation is consistent with our initial expectation.   

\section{Related Work}\label{relatedwork}

Text ranking has been an active area of research for decades~\cite{liu2009learning,DBLP:journals/corr/abs-2308-07107}.
One widely-adopted strategy is to rank candidate documents based on the relevance score between a query and the document~\cite{cossock2006subset}.
This strategy is referred to as the pointwise approach.
Additionally, pairwise~\cite{burges2005learning} and listwise~\cite{xia2008listwise} approaches, which consider the relative ordering of multiple documents in response to a query, have also gained great attention due to their strong performance.
In our work, we utilize the pointwise approach for efficient inference and the listwise approach for effective training,
making the best use of the both.

The precise calculation of the query-document relevance scores is the key to the pointwise inference, 
which has been generally dominated by supervised techniques~\cite{DBLP:conf/cikm/GuoFAC16}.
Initially, manually-crafted sparse features were employed to estimate these scores~\cite{DBLP:journals/jmlr/ChapelleC11}.
Subsequently, neural network models marked a turning point, showcasing their substantial promise~\cite{DBLP:journals/corr/PangLGXWC16}.
More recently, the progress of PLMs has led to remarkable advances in score calculation by using a pre-training and fine-tuning framework for text ranking~\cite{DBLP:conf/ecir/GaoDC21,DBLP:conf/sigir/JuYW21,DBLP:journals/corr/abs-2101-05667,zhang2023hybrid,li2023faa}.

As PLMs evolved into decoder-only LLMs, early work explored out-of-the-box strategies for text ranking to leverage the inherent strong reasoning capabilities of LLMs~\cite{sun2023chatgpt,qin2023large,ma2023zero,DBLP:conf/acl/ChoJSP23}.
For instance, one could simply ask LLMs to determine the relevance of a query-document pair, yielding a rudimentary solution~\cite{DBLP:journals/corr/abs-2211-09110,DBLP:journals/corr/abs-2310-14122,DBLP:conf/emnlp/Zhuang0KZ23}.
Subsequent works~\cite{DBLP:conf/emnlp/SachanLJAYPZ22,sgpt,drozdov2023parade}, propose the use of query generation likelihood based on a candidate document as a measure of relevance.
We follow this line of work for the relevance score definition.

Nevertheless, these out-of-the-box strategies often overlook the potential misalignment between LLMs and the specific requirements of text ranking tasks, which is a major issue that our work aims to address. 
Task-specific LLM training can help bridge this gap~\cite{sun2023instruction,rankllama,DBLP:journals/corr/abs-2312-02969}, as exemplified by RankLLaMA \cite{rankllama}, which uses the last token as the ranking basis and trains with ranking losses.

While dominant LLMs typically feature a decoder-only architecture, previous research on adapting encoder(-decoder) PLMs to text ranking remains highly pertinent~\cite{monobert,monot5,rankt5}. The underlying principles and core ideas of these studies underpin our approach, and their benefits apply across diverse model architectures.

\section{Conclusion}
In our work, we proposed a novel two-stage training paradigm to adapt LLMs to text ranking tasks.
Specifically, we first performed CPT on a large-scale weakly-supervised corpus to initially align LLMs with ranking objectives.
This is then followed by SFT on high-quality data along with full-query generation optimization and auxiliary objectives. 
Through this two-stage training paradigm, we achieved improved ranking performance on various LLMs in both in-domain and out-domain experimental settings. 
The significant gains exhibited by our approach highlight its effectiveness in enhancing the capabilities of LLMs for text ranking.
\section{Limitations}
While our two-stage adaptation can effectively enhance LLMs' text ranking capabilities, several limitations remain.
First, our CPT is actually independent of ranking objectives. Introducing ranking objectives similar to SFT during CPT worths further exploration.
Second, using a unified prompt for all ranking tasks might damage model generalization.
We generally refer to the texts of the BEIR benchmark as ``query'' and ``document'' in prompts, as shown in Eq.~\ref{eq:score}. 
In fact, these texts can be further classified.
Specifically, ``query'' involves ``title'', ``entity'', etc, and ``document'' involves ``argument'', ``news'', etc.
Finally, we evaluate our approach only on the BEIR benchmark. 
Testing it on diverse ranking tasks, such as the demonstration ranking of the in-context-learning scenario and knowledge ranking to mitigate hallucinations in LLMs, would better validate its broad applicability.

\section{Acknowledgement}
We sincerely thank the reviewers for their invaluable feedback, which significantly improved the quality of this work.
This work is supported by the National Natural Science Foundation of China (NSFC) Grant Nos.62336008 and Nos.62176180.

\bibliography{custom}

\clearpage
\appendix

\section{Appendix}
\label{sec:appendix}

\begin{figure}[h]
	\centering
	\includegraphics[width=0.35\textwidth]{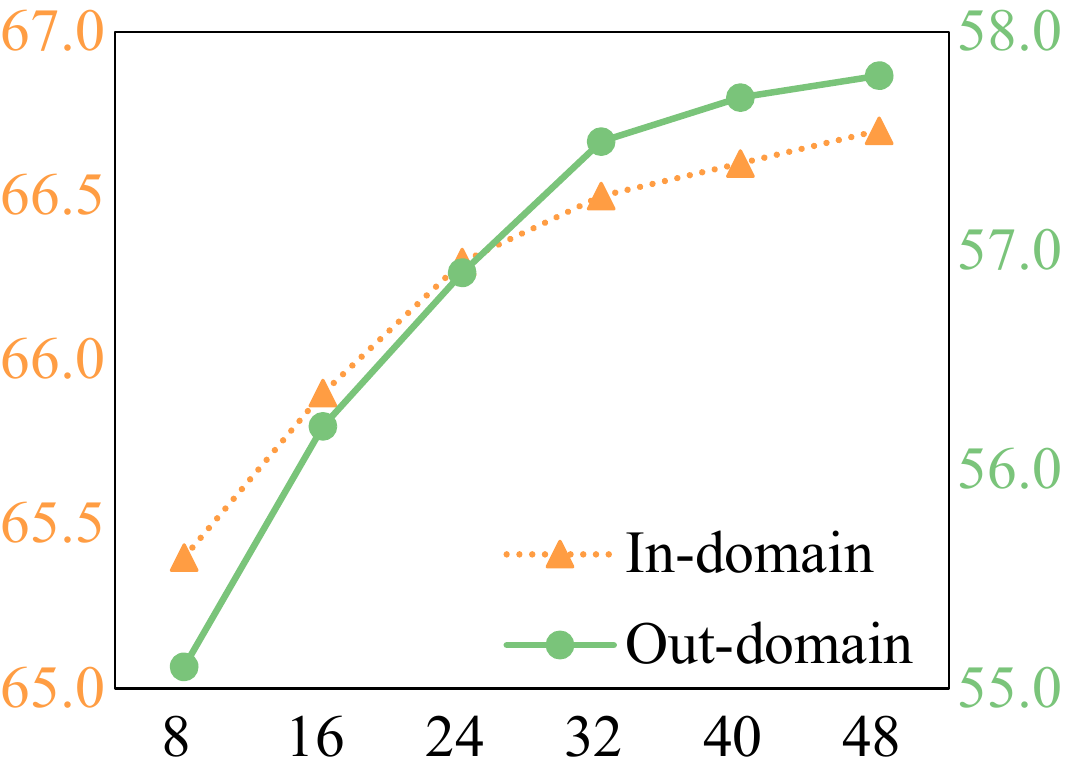}
	\caption{Results for various negative examples sizes $m$ in Eq.~\ref{eq:overall} on LLaMA-7B.}
	\label{fig:negsize}
\end{figure}

\paragraph{The Negative Examples Size.}
Negative example size $m$ in  Eq.~\ref{eq:overall} is the key to ranking loss.
We test the impact of $m$ on in-domain and out-domain performance in Figure~\ref{fig:negsize}.
Overall, increasing negative examples enhances ranking performance, consistent with traditional conclusions~\cite{rankt5,rankllama}.
Moreover, we find that out-domain performance is more sensitive to negative example size.
Specifically, as the $m$ increases from 8 to 48, the in-domain result increases by 1.3, while the corresponding increase in the out-domain result is 2.7.

\begin{figure}[h]
	\centering
	\includegraphics[width=0.35\textwidth]{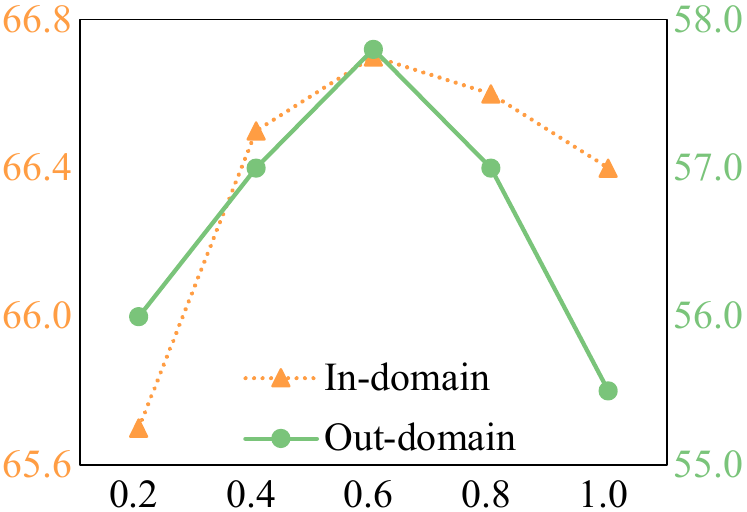}
	\caption{The effect of $\alpha$ in Eq.~\ref{eq:overall} on LLaMA-7B.}
	\label{fig:alpha}
\end{figure}

\paragraph{Balancing Ranking and Auxiliaries.}
During SFT, the trade-off between ranking and auxiliary objectives is controlled by $\alpha$ in Eq.~\ref{eq:overall}.
We analyze the impact of $\alpha$, as shown in Figure~\ref{fig:alpha}.
If the $\alpha$ is too low, the model would ignore the ranking objective, which is the key to SFT. 
Therefore, when the $\alpha$ is 0.2, the model performs poorly in both scenarios.
Conversely, high $\alpha$ values barely affect in-domain results.
From $\alpha$=0.6 to 1.0, the in-domain performance only decreases by 0.3. 
However the corresponding out-domain result drops significantly by 2.3.
This is because over-emphasis on the ranking objective leads to overfitting in the in-domain scenario and affects model generalization.

\end{document}